\documentclass[twocolumn,floatfix,aps,showpacs]{revtex4}
\usepackage{flafter}
\usepackage{color}
\usepackage{braket}
\usepackage{amsmath}
\usepackage{amssymb}
\usepackage{graphicx}
\usepackage{dcolumn}
\usepackage{bm}
\usepackage{epsfig}
\usepackage{subfigure}

\begin{document}

\title{Hydrodynamic theory of motion of quantized vortex rings in trapped
superfluid gases.}
\author{Lev P. Pitaevskii$^{1,2}$}
\date{\today }

\begin{abstract}
I study vortex ring oscillations in a superfluid, trapped in an elongated
trap, under the conditions of the Local Density Approximation. On the basis
of the Hamiltonian formalism I develop a hydrodynamic theory, which is
valid for an arbitrary superfluid and depends only on the equation of state.
The problem is reduced to an ordinary differential equation for the ring radius.
The cases of the dilute BEC  and the Fermi gas at unitarity are investigated
in detail. Simple analytical equations for the periods of small oscillations are
obtained and the equations of non-linear dynamics are solved in quadratures. The
results agree with available numerical calculations. Experimental
possibilities to check the predictions are discussed.
\end{abstract}

\affiliation{1 Dipartimento di Fisica, Universit\`{a} di Trento and INO-CNR BEC Center,
I-38123 Povo, Italy\\
2 Kapitza Institute for Physical Problems RAS, Kosygina 2, 119334 Moscow,
Russia}
\maketitle

\textit{Introduction.} The quantized vortex ring in a superfluid is one of
the most unusual objects of modern physics. It can be quite macroscopic in
size but still keeps its quantum nature, carrying one quanta of circulation.
Rings of large size have small phase velocity and play a crucial role in the
phenomenon of critical velocity. The existence of such rings was predicted
by Feynman~\cite{F55}. They were discovered, in an indirect way, in
experiments on ion motion in liquid $^{4}$He~\cite{Ray64}. Vortex rings were
also observed in gaseous Bose-Einstein Condensates (BEC) in traps  \cite%
{Corn01}. There is every reason to believe that the development of
experimental technique will permit the detailed investigation of rings, both
in trapped BEC and Fermi gases near unitarity. The dynamics of rings
under such conditions should be quite peculiar. In a uniform fluid a ring
always moves with an anomalous energy-velocity relation - its velocity
decreases with increasing energy. Under the non uniform conditions in a trap
the situation is different. It was discovered in \cite{J99} that in a
spherical trap a configuration of maximum energy exists, where a circular
ring is at rest. It was checked in \cite{G02} that the same situation
takes place also in an elongated trap. An initial deviation from this
equilibrium configuration will result in oscillations of the ring. These
oscillations in a spherical trap were investigated numerically in \cite{J99}.

The problem of oscillation of a ring along a symmetry axis of a superfluid
sample in an elongated harmonic trap is one of the most natural subjects of experimental
investigation. Numerical simulations have been recently performed in   \cite{B13}
for a Fermi gas at unitarity  using a
time-dependent density functional theory and in \cite{RM13} for a BEC using
 the Gross-Pitaevskii (GP) equation. The main goal of these papers is the
interpretation of experimental results \cite{Z13}. An approximate theory
for ring motion in a cylindrical trap in the presence of dissipation was
developed in \cite{FSh99}.

\textit{General theory.} In this letter I will present the Hamiltonian
theory of the motion of a ring of radius $R$ along the axis of a superfluid
gas in a harmonic trap. I assume that the trap is elongated, $\omega _{\perp
}\gg \omega _{z}$. I also assume the Local Density Approximation (LDA)
conditions, that is that the transverse radius of the gas is much larger than 
the healing length, $R_{\perp }\gg \xi 
$. Moreover I will 
assume that these conditions are satisfied strongly in the sense that also
the logarithmic factor $L\equiv \log (R/\xi )$, which enters in the theory,
is large and can be considered as a constant in all calculations. We will
see that in this \textquotedblleft logarithmic\textquotedblright
approximation one can solve the problem analytically in a simple way for a
superfluid of any nature. I will finally assume, that $R \sim R_{\perp}$,
excluding very small rings.

The LDA conditions permit us to use hydrodynamics. The energy of a ring, i.
e. the Hamiltonian, can be written directly. The key point is that the energy
can be obtained by integrating the kinetic energy of the flow $\rho 
\mathbf{v}^{2}/2$ over the volume ($\rho $ is the density of the fluid).
In the logarithmic approximation the main contribution to the
integral is due to a small region near the vortex line. Thus one can use the
expression for the energy in a uniform fluid, taking the density $\rho (r,z) $
to be its value near the ring:%
\begin{equation}
E_{R}(R,Z)=\frac{2\pi ^{2}\hbar ^{2}}{M^{2}}R\rho \left( R,Z\right) \log
\left( \frac{R}{\xi }\right) \;,  \label{ER}
\end{equation}%
where $M$ is the atomic mass $m$ for the Bose superfluid and the pair
mass $2m$ for the Fermi one, and $Z$ is the $z$-coordinate of the
ring. Notice that the coordinate dependence of the density in the LDA regime
can be written in the form%
\begin{equation}
\rho \left( r,z\right) =\rho \left[ \mu \left( 1-r^{2}/R_{\perp
}^{2}-z^{2}/R_{z}^{2}\right) \right] ,  \label{rho}
\end{equation}%
where $\mu $ is the chemical potential in the center of the trap and $%
R_{\perp }=\left( 2\mu /m\omega _{\perp }^{2}\right) ^{1/2}$, $R_{z}=\left(
2\mu /m\omega _{z}^{2}\right) ^{1/2}$are the Thomas-Fermi (TF) radii of the
fluid.

The momentum of the ring can be calculated as $P_{R}(R,Z)=\left( \hbar
/M\right) \int \rho \left( r,z\right) \partial _{z}\phi d^{3}r$, where $\phi 
$ is the phase of the order parameter (condensate wave function in BEC
case). However, for an elongated trap with $R_{z}\gg R_{\perp }$ one can
substitute $z\approx Z.$ Then the integration will be reduced to an
integration on the surface, stretched on the ring aperture, where the phase $%
\phi $ undergoes a $2\pi $ jump:%
\begin{equation}
P_{R}\approx \frac{\hbar }{M}\int \rho \left( r,Z\right) \partial _{z}\phi
d^{3}r=\frac{2\pi \hbar }{M}\int_{0}^{R}\rho \left( r,Z\right) 2\pi rdr.
\label{PR}
\end{equation}%
The velocity of the ring can be calculated with the Hamilton equation as 
\begin{equation}
V\equiv V_{z}=\left( \frac{\partial E_{R}}{\partial P_{R}}\right) _{Z}=\frac{%
\left( \partial E_{R}/\partial R\right) _{Z}}{\left( \partial P_{R}/\partial
R\right) _{Z}}.  \label{V}
\end{equation}%
Such a method was used in \cite{PS} to calculate o the velocity for a
uniform fluid. The possibility to use this equation in a non-uniform fluid,
if the energy and momentum are calculated properly, is a central point of
the theory. Taking into account that, according to (\ref{PR}), $\left(
\partial P_{R}/\partial R\right) _{Z}=\left( 4\pi ^{2}\hbar /M\right) R\rho
\left( R,Z\right) $, we finally obtain an elegant equation ($\rho =\rho
\left( R,Z\right) )$: 
\begin{equation}
V(R,Z)=\frac{\hbar }{2M}\frac{L}{R\rho }\left( \frac{\partial \left( R\rho
\right) }{\partial R}\right) _{Z}\;.  \label{VR}
\end{equation}%
Notice that the equation (\ref{VR}) admits a transparent interpretation. The
quantity $F=\frac{1}{2\pi R}\left( \frac{\partial E_{R}}{\partial R}\right) $
is the force, acting on a unit of length of the ring. According to the
Magnus equation, the ring drifts then with the velocity $FM/\left( 2\pi
\hbar \rho \right) $ in agreement with (\ref{VR}).

Equations (\ref{ER}) and (\ref{VR}) give a full description of the motion of
the ring in a trap. Notice that the theory is completely hydrodynamic in its
nature. Properties of the fluid enter only through the equation of state $%
\rho \left( \mu \right) $. It is convenient to introduce the dimensionless
variables $X=Z/R_{z}$ and $Y=R/R_{\perp }$. Energy can be presented as $%
E_{R}(R,Z)=\left( 2\pi \hbar ^{2}R_{\perp }\rho _{0}L/M^{2}\right) f(Y,X)$,
where $\rho _{0}$ is the density in the center of the trap and $f$ is a
dimensionless function. The trajectory of the ring on the $X,Y$ plane is
given by the energy conservation equation%
\begin{equation}
f\left( Y,X\right) =f_{0}\ ,  \label{f0}
\end{equation}%
where $f_{0}$ fixes the energy of the ring. The plane where the ring can be
at rest is always at $X=0$. The equilibrium radius $R_{EQ}=R_{\perp }Y_{EQ}$
is defined by the equation $\partial f(Y,0)/\partial Y=0.$ The energy $E_{R}$
has a maximum at this point. 
The equation $f\left( Y,0\right) =f_{0}$ has two positive solutions, $Y_{0}$
and $Y_{1}$, where $0<Y_{0}<Y_{1}$. Then $R_{\min }=R_{\perp }Y_{0}$ is the
minimal radius of the ring on the given trajectory and $R_{\max }=R_{\perp
}Y_{1}$ is the maximal one. Below I will consider $Y_{0}$ as an "initial
point" of the trajectory. The equation $\left( \partial f/\partial Y\right)
_{X}=0$ defines the line of "turning points" on $Y,X$ plane, where the
velocity changes sign. Together with (\ref{f0}) it gives the turning point $%
Y_{A},X_{A}$ for a trajectory of given energy. It follows that a ring with $%
Y<Y_{A}$ moves in the same direction as in an uniform fluid and a ring with $%
Y>Y_{A}$ in the opposite direction. The amplitude of the oscillations in $z$%
-direction is $Z_{A}=R_{z}\left\vert X_{A}\right\vert $. It is worth noting,
that the ring at rest has zero velocity, but finite momentum, in  analogy
with rotons in superfluid $^{4}$He. 
\begin{figure}[t]
\includegraphics[width=0.98\columnwidth]{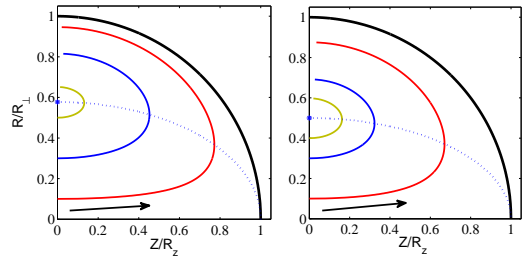}
\caption{(Color online) Trajectories of a vortex ring in the $R,Z$ plane.
The left panel corresponds to BEC, the right one - to a Fermi gas at
unitarity. Different curves correspond to different values of minimal radius 
$Y_{0}=R_{min}/R_{\perp }$, which can be seen on the $Y$ axis. The bold
lines (black online), connecting the points (0,1) and (1,0), are the TF
boundary of the fluid. The dashed line (blue online) connects the turning
points of the trajectories. The stars on the $Y$ axises show the equilibrium
values of the radius.}
\label{Fig1}
\end{figure}
The quantity, which can be easily measured, is a period of oscillation of
the ring. One can calculate  the period of small
oscillations $T_{0}$ in a general form by writing the energy
near the point $R=R_{EQ}$ and $Z=0$ in the oscillator form $E_{R}(R,Z)-E_{R}\left(
R_{EQ},0\right) \propto -\left[ \left( \frac{T_{0}}{2\pi }\right)
^{2}V^{2}+Z^{2}\right] $. Direct calculation gives%
\begin{equation}
T_{0}=2\sqrt{2}T_{z}\frac{M\mu }{mL\hbar \omega _{\perp }}\left( \frac{f^{2}%
}{\partial^{2}_{X}f\partial^{2}_{Y}f}\right) ^{1/2}\;,  \label{T0}
\end{equation}%
where $T_{z}=2\pi /\omega _{z}$ is the trap period  and quantities in the
parenthesis should be taken at $X=0,Y=Y_{EQ}$. In LDA regime $(\mu /L\hbar
\omega _{\perp })\gg 1$ and $T_{0}\gg T_{z}$, as revealed in numerical
calculations in \cite{B13},\cite{RM13}. Notice that the prefactor, fixed by
the parameters of the system, can be presented as $\left( T_{z}M\mu /mL\hbar
\omega _{\perp }\right) =$ $\left( \pi MR_{\perp }R_{z}/L\hbar \right) $,
demonstrating a simple dependence on pure geometric factors. 
\begin{figure}[t]
\includegraphics[width=0.98\columnwidth]{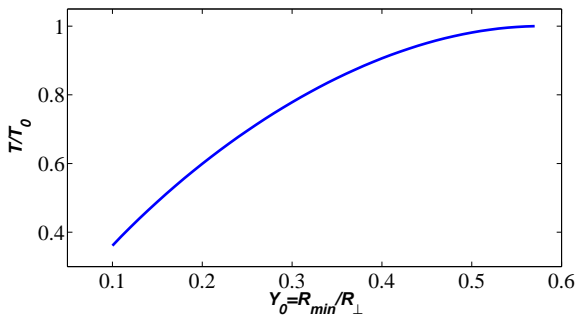}
\caption{(Color online) The period of oscillations in the units of the
period of small oscillations as a function of the minimal radius $Y_{0}$ for
a ring in BEC. }
\label{Fig1A}
\end{figure}

One can find the time dependence of $Z$ and $R$ from the equation $t=\int 
\frac{dZ}{V}$, where the integral should be taken along the trajectory. It
is more convenient to go to variable $R$. Using Eq. (\ref{VR}) we get%
\begin{equation}
t\left( Y\right) =-T_{z}\frac{M\mu }{\pi ^{2}mL\hbar \omega _{\perp }}2\pi
f_{0}\int_{Y_{0}}^{Y}\left( \frac{\partial X}{\partial f}\right) _{Y}dY\ .
\label{t}
\end{equation}%
The period for oscillations for an arbitrary amplitude can be found as $%
T=2t(Y_{1})$. This quantity is interesting, because it reflects the peculiarity
of the
dynamics of the rings, and also important, because it is difficult to
observe oscillations of a small amplitude in an experiment.

\textit{Vortex ring in a trapped dilute BEC.} In a dilute BEC the
chemical potential is $\mu =gn$, where $g$ is the coupling constant and $n$ is
the atom density. This means that the energy function $f$ \ is%
\begin{equation}
f(Y,X)=Y\left( 1-Y^{2}-X^{2}\right) \ .  \label{fB}
\end{equation}%
Then the equilibrium radius is $R_{EQ}=R_{\perp }Y_{EQ}=R_{\perp }/\sqrt{3}.$
This value coincides with one obtained in \cite{J99} in the logarithmic
approximation for a spherical trap. The line of the turning points is $%
3Y_{A}+X_{A}^{2}=1$. The minimal radius of the ring is related to the energy
as $f_{0}=Y_{0}\left( 1-Y_{0}^{2}\right) $ and the maximal radius $Y_{1}=%
\frac{1}{2}\sqrt{4-3Y_{0}^{2}}-\frac{1}{2}Y_{0}$. The equation of the
trajectory with initial radius $Y_{0}$ can be written as $Y\left(
1-Y^{2}-X^{2}\right) =Y_{0}\left( 1-Y_{0}^{2}\right) $ and the amplitude of
oscillation can be expressed through energy as $X_{A}=\sqrt{1-\frac{3}{%
2^{2/3}}f_{0}^{2/3}}$. Trajectories for different values of $Y_{0}$ are
shown in the left panel in Fig. 1. (One can see values of $Y_{0}$ as initial
values of $Y$ on the $y-$axis,)

The period of small oscillations can be calculated according to Eq. (\ref{T0}%
). A simple calculation gives 
\begin{equation}
\frac{T_{0}^{\left( B\right) }}{T_{z}}=\frac{4}{3\sqrt{3}}\frac{\mu }{L\hbar
\omega _{\perp }}\approx 0.77\frac{\mu }{L\hbar \omega _{\perp }}\ .
\label{T0B}
\end{equation}%
Scaling of $T$ as $gn/\log \left( gn\right) $ was predicted in \cite{RM13}
on the basis of qualitative considerations. For oscillations of an arbitrary
amplitude let us present the period as $T^{\left( B\right) }=T_{z}\left( \mu
/\pi ^{2}L\hbar \omega _{\perp }\right) \tau ^{\left( B\right) }$. One
obtains the expression for the dimensionless period $\tau $ 
\begin{equation}
\tau ^{\left( B\right) }=\int_{Y_{0}}^{Y_{1}}\frac{2\pi f_{0}dY}{\sqrt{%
Y(Y-Y_{0})(Y_{1}-Y)(Y-Y_{2})}}\ ,  \label{tauBint}
\end{equation}%
where $Y_{2}=-\frac{1}{2}\sqrt{4-3Y_{0}^{2}}-\frac{1}{2}Y_{0}$ is a negative
root of the equation $f\left( Y,0\right) =f_{0}$. Equation (\ref{tauBint})
can be expressed in terms of the complete elliptic integral of the first
order:%
\begin{equation}
\tau ^{\left( B\right) }=\frac{4\pi f_{0}}{\sqrt{Y_{1}\left(
Y_{0}-Y_{2}\right) }}K(k),k=\sqrt{\frac{\left( Y_{1}-Y_{0}\right) \left(
-Y_{2}\right) }{Y_{1}\left( Y_{0}-Y_{2}\right) }}.  \label{tauBell}
\end{equation}%
\begin{figure}[t]
\includegraphics[width=0.98\columnwidth]{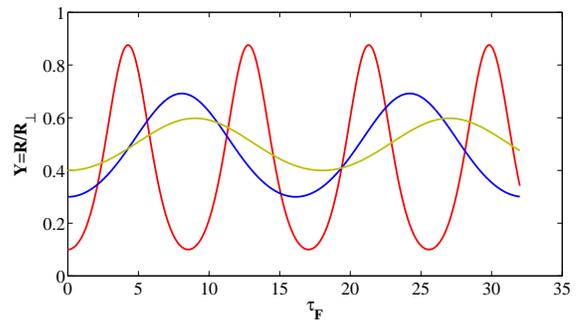}
\caption{(Color online) The time dependence of the radius of a ring,
oscillating in the Fermi gas at unitarity. Different curves correspond to
different values of minimal radius $Y_{0}=R_{min}/R_{\perp }$, which can be
seen on the $Y$ axis. }
\label{Fig2}
\end{figure}
For small oscillations $Y_{0}\rightarrow Y_{1}\rightarrow 1/\sqrt{3}%
,f_{0}=2/(3\sqrt{3}),Y_{2}=-2/\sqrt{3}.$ Then $\tau ^{\left( B\right)
}\rightarrow 4\pi ^{2}/3\sqrt{3}$ in accordance with (\ref{T0B}). At small $%
Y_{0}$ one gets formally $\tau ^{\left( B\right) }\approx 2\pi Y_{0}\ln
(16/Y_{0})$. However, this limit violates the applicability of the
approximation. 
\begin{figure}[t]
\includegraphics[width=0.98\columnwidth]{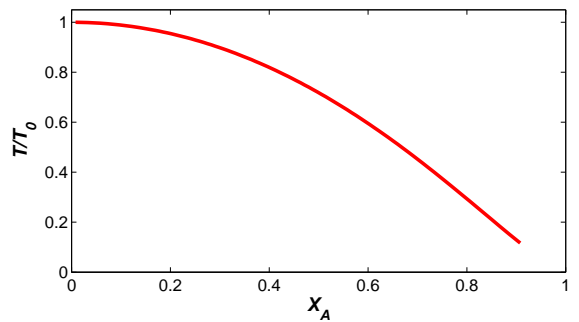}
\caption{(Color online) The period of oscillations in the units of the
period of small oscillations as a function of the amplitude of oscillations $%
X_{A}$ for a ring in the Fermi gas at unitarity. }
\label{Fig3}
\end{figure}
The dependence of the period on the minimal radius of a ring $Y_{0}$ is
shown in Fig. 2.

\textit{Vortex ring in a trapped Fermi gas at unitarity.} In a Fermi gas at
unitarity $\rho \propto \mu ^{3/2}$. Correspondingly, the energy function is 
\begin{equation}
f(Y,X)=Y(1-Y^{2}-X^{2})^{3/2}\ .  \label{fF}
\end{equation}%
Equation (\ref{f0}) then defines the trajectories in the $X,Y$ plane, which
are presented in the right panel of Fig. 1. The equilibrium radius is now $%
Y_{EQ}=1/2$ and the period of small oscillations can be found immediately
according to (\ref{T0}):%
\begin{equation}
\frac{T_{0}^{\left( F\right) }}{T_{z}}=\frac{\sqrt{3}}{\sqrt{2}}\frac{\mu }{%
L\hbar \omega _{\perp }}=\allowbreak \allowbreak 1.22\frac{%
\mu }{L\hbar \omega _{\perp }}\ .  \label{TF0}
\end{equation}%
The calculations for arbitrary amplitudes are more cumbersome that in the
BEC case and I will present them in short. The equation for the line of the
turning points is $4X_{A}^{2}+Y_{A}^{2}=1$. The maximal value of $X_{A}$, i.
e. the amplitude of oscillations, is $X_{A}=\sqrt{1-\frac{4}{3^{3/4}}%
f_{0}^{1/2}}$ and the corresponding $Y_{A}=\frac{1}{\left( 3\right) ^{3/8}}%
f_{0}^{1/4}$. Minimal and maximal values of the radius are given by the
equation $f(Y,0)=f_{0}$, which, introducing the variable $y=Y^{2/3}$, can
be transformed to $y(1-y^{3})=f_{0}$. For a trajectory with the minimal
radius $Y_{0}=y_{0}^{3/2}$ this equation can be presented as $\left(
y-y_{0}\right) Q(y)=0$, where 
\begin{equation}
Q(y)=y^{3}+y^{2}y_{0}+yy_{0}^{2}+y_{0}^{3}-1\ .
\end{equation}%
The equation $Q(y)=0$ has 3 roots. The root $y_{1}$ is real and defines the
maximum value of the radius, $Y_{1}=y_{1}^{3/2}$. Roots $y_{2}$ and $y_{3}$
are complex conjugated. One can find the roots analytically or numerically.

Changing the variable of integration in (\ref{t}) from $Y$ to $y$, I present
the time of motion as $t^{\left( F\right) }=T_{z}\left( 2\mu /\pi ^{2}L\hbar
\omega _{\perp }\right) \tau ^{\left( F\right) }$, where 
\begin{equation}
\tau ^{\left( F\right) }(y)=\int_{y_{0}}^{y}\frac{2\pi f_{0}^{2/3}dy}{\sqrt{%
(y_{1}-y)\left( y-y_{0}\right) \left( y^{2}-2\mathrm{Re}(y_{2})y+\left\vert
y_{2}\right\vert ^{2}\right) }}\ .  \label{tF}
\end{equation}%
The integral can be expressed in terms of an elliptic integral (see \cite{GR}%
, Eq. 3.145)%
\begin{equation}
\tau ^{\left( F\right) }=\frac{4\pi f_{0}^{2/3}}{\sqrt{pq}}F\left( \varphi
,k\right) \ ,
\end{equation}%
where the parameters are%
\begin{equation}
\varphi =2\mathrm{arccot}\sqrt{\frac{q\left( y_{1}-y\right) }{p\left(
y-y_{0}\right) }},k=\sqrt{\frac{\left( y_{1}-y_{0}\right) ^{2}-\left(
p-q\right) ^{2}}{4pq}}
\end{equation}%
and 
\begin{eqnarray}
p &=&\left[ \left\vert y_{0}\right\vert ^{2}+\left\vert y_{2}\right\vert
^{2}-2y_{0}\mathrm{Re}(y_{2})\right] ^{1/2},  \notag \\
q &=&\left[ \left\vert y_{1}\right\vert ^{2}+\left\vert y_{2}\right\vert
^{2}-2y_{1}\mathrm{Re}(y_{2})\right] ^{1/2}.
\end{eqnarray}%
Inverse dependence $y(\tau ^{F})$ can be expressed in terms of the Jacobi cn(%
$t,k)$ function. The period of oscillations is $2t^{\left( F\right) }(Y_{1})$%
. For small amplitudes the result coincides with (\ref{TF0}). For a ring of
small radius, $Y_{0}\rightarrow 0$, one gets $\tau ^{\left( F\right)
}\rightarrow 15.3Y_{0}^{2/3}$. Notice that the authors of \cite{B13} used a
slightly different dependence of period on radius for relatively small
rings: $T^{\left( F\right) }\propto R$ as opposite to $T^{\left( F\right)
}\propto R^{2/3}$ here. In Fig. 3 I show the time dependence of the
radius of a ring, oscillating in the Fermi gas at unitarity. One can see
that the oscillations are almost harmonic. Only for the small initial
radius $Y_{0}=0.1$  the difference between growing and shrinking motion,
which was revealed in \cite{RM13}, can be noticed. In contrast, the dependence
of the period on the amplitude of oscillations is strong. This dependence is
shown in Fig. 4. It is interesting, that the analogous curve for a ring in 
BEC is practically indistinguishable from Fig. 4.

The results of the present analytical theory are in qualitative agreement
with numerical calculations \cite{B13,RM13}: the period of
oscillations is much longer than $T_{z}$ and increases when the radius
of the ring decreases or the interaction increases. I tried to compare
quantitatively my results with numerical calculations \cite{RM13}. These
calculations were produced for a ring in a BEC with $R_{\perp }/\xi =2\mu
/\hbar \omega _{\perp }\approx 27$, giving $L\sim 3.3$ and $\omega _{\perp
}/\omega _{z}=4$, that, of course, does not ensure applicability of my
asymptotic theory. However, formal use of the theory gives $T^{\left(
B\right) }/T_{z}\approx 3$ for $R_{\min }/a_{z}\approx 1.2$. The data
presented in Fig. 5 of \cite{RM13} give $T^{\left(B\right) }/T_{z}\approx
2.6 $ in surprisingly good agreement. Quantitative comparison 
with calculations of Ref. \cite{B13} for the Fermi gas at unitarity is unreasonable,
because the LDA conditions are not satisfied there.

It is worth noting that the present theoretical scheme can be generalized
to relax the strong inequality, which I used. For example, one can calculate
the energy and momentum numerically, as done in \cite{G02}, and still use
equation (\ref{V}). Of course, there is no problem to use this theory for a
non-harmonic trap. The theory also can be generalized for more complicated
"solitonic vortices" excitations, observed in numerical calculations \cite%
{BR02,KP02,KP03}.  An interesting direction of applications of
the theory is the dynamics of topologically nontrivial excitations in two
interpenetrating superfluids. (See, for example, \cite{SS02,U04}.)

Experimental confirmations of the present results demand strong LDA
conditions. However, I believe that in the Fermi gas they can be satisfied
in a natural way. For example, in experiments \cite{Z12} the value $%
R_{\perp}/\xi \approx 100$ of the LDA parameter at $\omega_{\perp}/%
\omega_{z}=6.2$ was reached, which is sufficient for the theory, and nothing
prevents using even larger values. The problem of the creation of a ring in
a controlled way is not an easy one. It can be solved as a result of the
snake instability of a soliton, as it already observed in \cite{Corn01}. It
was suggested in \cite{J98} to create a ring by a moving bright spot of
laser light, in analogy with rings creation by moving impurities in
experiments \cite{Ray64}. In experiments \cite{Kohl09} and \cite{Massimo11}
gravitational accelerations of impurity atoms in a trapped superfluid was
observed. This effect can also be used to create rings. Recently, an
ingenious method of creating vortex rings in  BEC  by a fast increase
interaction near the Feshbach resonance was suggested in \cite{Berl13}. The
same method can be used in a Fermi superfliud.

To conclude, I have developed a Hamiltonian theory for the radius and center
of a quantized vortex ring, oscillating along an axis of a superfluid,
trapped in an elongated cylindrical trap. The theory is valid in the strong
Local Density Approximation, when $L=\log \left( R_{\perp }/\xi \right) \gg
1 $. The theory is pure hydrodynamical and demands only knowledge of the
equation of state of the fluid. It occurs that the equations can be solved
in quadratures and the period of oscillations is scaled as $T\sim
T_{z}\left( \mu /L\hbar \right) $. For the cases of  dilute BEC  and
Fermi gas at unitarity simple expressions for the periods are obtained and
solutions can be expressed in terms of elliptic integrals. Possible
generalizations are mentioned. Ways of verifying of the predictions in experiments
are discussed in short.

I thank D.~Papoular for useful advice, M.~McNeil~Forbes, M.~Reichl, and M.~Zwierlein for
additional information on their papers \cite{B13,RM13,Z12} and F.~Dalfovo and
R.~Scott for critical reading of the manuscript.

The work was supported by the ERC through the QGBE grant, the Provincia
Autonoma di Trento, and the Italian MIUR through the PRIN-2009 grant.

\end{document}